\documentclass[12 pt, a4paper]{article}
\usepackage[utf8]{inputenc}
\usepackage{jheppub}
\pdfoutput=1
\usepackage{amsmath}
\usepackage{physics}
\usepackage{graphicx}
\usepackage{subcaption}
\usepackage{xcolor}
\usepackage{multirow}
\newcommand{\Lagr}{\mathcal{L}}
\makeatletter

\def\lsim{\compoundrel<\over\sim}
\def\compoundrel#1\over#2{\mathpalette\compoundreL{{#1}\over{#2}}}
\def\compoundreL#1#2{\compoundREL#1#2}
\def\compoundREL#1#2\over#3{\mathrel
         {\vcenter{\hbox{$\m@th\buildrel{#1#2}\over{#1#3}$}}}}
\makeatother

\title{\vspace{-2cm} 
 \begin{flushright}
  {\normalsize INR-TH-2021-023}
 \end{flushright}
 \vspace{0.5cm} Gravitational waves from first-order electroweak phase transition in a model with light sgoldstinos}

\author[a,b]{S. Demidov,}
\author[a,b]{D. Gorbunov}
\author[a,c]{and E. Kriukova}

\affiliation[a]{Institute for Nuclear Research of the Russian Academy of Sciences, \\
  60th October Anniversary pr-ct 7a, Moscow 117312, Russia}
\affiliation[b]{Landau Phystech School of Physics and Research, Moscow Institute of Physics and Technology, \\
	Institutskiy per. 9, Dolgoprudny 141700, Russia}
\affiliation[c]{Faculty of Physics, Lomonosov Moscow State University,\\
  Leninskiye Gory 1-2, Moscow 119991, Russia}
  
\emailAdd{demidov@inr.ac.ru}
\emailAdd{gorby@inr.ac.ru}
\emailAdd{kryukova.ea15@physics.msu.ru}

\abstract{ We study previously unexplored possibility of triggering the first order electroweak phase transition (EWPT) by interactions of the Standard Model (SM) particles with the sector responsible for low scale supersymmetry breaking. The low-energy theory
 apart from the SM particles contains additional scalar degrees of freedom --- sgoldstinos --- which
  contribute to the effective scalar potential and thus can trigger the first order EWPT. 
   Remarkably, the latter requires only moderate couplings in the scalar sector. The perturbative description in terms of the effective theory seems formally to be applicable upto the scale of supersymmetry breaking: the Landau pole in the scalar sector is above $10^8$-$10^9$\,GeV.  
  We calculate the gravitational wave signal generated at this transition (it can be tested, e.g. by LISA, BBO and DECIGO) and briefly discuss the collider phenomenology of this scenario.}

\arxivnumber{2112.06083}

\begin{document}
\maketitle

\section{Introduction}
There are plenty of well motivated new physics scenarios which
were proposed to solve phenomenological problems of
the Standard Model (SM), and
supersymmetry~\cite{Nilles:1983ge,Haber:1984rc} is among the most 
attractive paradigms for the SM extension. Technically natural
supersymmetric scenarios predict superpartners with masses near
the electroweak scale. Yet, so far no signals from supersymmetry have been seen at
LHC experiments~\cite{atlas,cms} except for several
existing anomalies in the experimental data which are still
inconclusive. 

Surely, in a realistic supersymmetric scenario the supersymmetry
must be broken. In the case of spontaneously broken supersymmetry
the model should contain the sector which is responsible for this
spontaneous supersymmetry breaking (SSB). In the framework of global supersymmetry
the SSB implies existence of the Goldstone fermion, 
{\it goldstino}~\cite{Volkov:1973ix}, which becomes the longitudinal component of
gravitino in supergravity (local supersymmetry)  extensions~\cite{Cremmer:1978iv}. In a full
model goldstino should be a member of supermultiplet from the hidden sector. It is phenomenologically possible that some of
particles inhabiting this sector are lighter than 
superpartners of the SM particles and in some classes of models they may show up in
experiments before the superpartners. One of such scenarios, which we
consider in this study, is the models with low scale supersymmetry
breaking that can appear in several
setups~\cite{Ellis:1984kd,Ellis:1984xe,Giudice:1998bp,Dubovsky:1999xc}. In 
the simplest case the low-energy part of the hidden sector in these models consists
of a single chiral superfield 
\begin{equation}
\label{chiral}
\Phi = \phi + \sqrt{2}\psi\theta +
\theta\theta F_\Phi\,, 
\end{equation}
where $\psi$ is goldstino, $\phi=\frac{1}{\sqrt{2}}(s+ip)$ 
are its scalar superpartners, {\it sgoldstinos}, and $F_\Phi$ is an auxiliary
field which upon SSB gains non-zero vacuum expectation value
$\langle F_\Phi\rangle = F$ with $\sqrt{F}$ being the energy scale
of supersymmetry breaking. Within the low scale supersymmetry, it is assumed
that this scale is not far from the electroweak scale. Correspondingly,
the gravitino has the mass $m_{3/2}\sim\frac{F}{M_{\rm Pl}}$ and
this state is the lightest supersymmetric particle. Sgoldstinos,
on the other hand, acquire non-zero masses through interactions in
the hidden sector (sgoldstinos are massless at the tree level of perturbation theory). In this study we consider the scenario with 
sgoldstinos at the electroweak (or subTeV) scale, while other
superpartners are significantly heavier. General aspects of this setup 
were discussed in refs.~\cite{Brignole:1996fn,Brignole:2003cm} 
and this scenario exhibits quite reach phenomenology, see,
e.g.~\cite{Brignole:2000wd,Perazzi:2000id,Perazzi:2000ty,Gorbunov:2000cz,Gorbunov:2000ht,Gorbunov:2002er,Demidov:2004qt,Dudas:2012fa,Demidov:2016gmr}. In 
particular, presence of relatively light sgoldstinos can noticeably affect the Higgs sector of the model, see 
e.g.~\cite{Petersson:2011in,Astapov:2014mea,Demidov:2020jne}.   

Here we consider yet unexplored possibility that new light scalar
degrees of freedom, from the sector where supersymmetry gets broken, 
contribute to the low-energy effective potential and can trigger the
first order electroweak phase transition. In the decoupling regime the
low-energy theory contains additional complex-valued scalar field
whose coupling constants to the SM fields are determined mainly by
ratios of soft supersymmetry  breaking parameters to the supersymmetry breaking
scale squared. Possibility of the first order EWPT in the SM extensions 
with new scalars was studied in literature, see e.g.
refs.~\cite{Profumo:2007wc,Espinosa:2011ax,Cline:2012hg,Profumo:2014opa,Jiang:2015cwa,Kurup:2017dzf,Demidov:2017lzf,Gould:2019qek}. 
The interest to the first order EWPT is associated with possible
realisation of the electroweak 
baryogenesis~\cite{Kuzmin:1985mm,Morrissey:2012db} and 
generation of observable gravitational wave
signal~\cite{Kamionkowski:1993fg,Apreda:2001us,
  Grojean:2006bp,Huber:2008hg,Caprini:2018mtu}. 
In this study we concentrate on the latter option. The gravitational wave signal from the first-order phase transition due to SUSY breaking in the hidden sector was previously studied in \cite{Craig:2020jfv}. We, however, scrutinize on possible influence of the sector responsible for SUSY breaking on EWPT. 

We study a
part of the parameter space of supersymmetric model with light
sgoldstinos which admits the first order EWPT in the early Universe. 
Naturally it is provided by interactions between the new ingredients in the scalar sector and the SM Higgs field. Typically, for the first order EWPT to take place, the corresponding couplings must be rather large, see model examples in the references above, which exhibit Landau poles not far from the electroweak scale. It requires new physics there and rises the question of whether the description of the EWPT in terms of the effective low-energy theory is fully justified. In our case the couplings in the scalar sectors turn out to be rather moderate, and the corresponding Landau poles lifted up to the energy scale as high as $10^8$-$10^9$\,GeV. Therefore one can formally use the effective description of this model upto this scale where it is completed by introducing new particles: those responsible for the supersymmetry breaking in the hidden sector, those responsible for the mediation of the supersymmetry breaking to the visible (SM) sectors, the SM superpartners etc. Evidently, the supersymmetry breaking scale may be well lower, moreover, one can expect the SM superpartners mass scale to be lower, and so the model is modified at the lower scale of their masses. In any case the effective description works in the fairly wide energy region allowing for a broad class of supersymmetric models to complete it at high energies. Therefore, applying the effective model description  
we calculate the gravitational wave signal that can be observed at LISA and 
proposed experiments BBO, DECIGO and Ultimate
DECIGO~\cite{Schmitz:2020syl}. We also discuss a promising collider
phenomenology pertinent to this scenario.

The rest of the paper is organised as follows. In Section~\ref{SUSYmodel}
we introduce the supersymmetric model with low-energy goldstino
supermultiplet. Section~\ref{Sec:potential}
describes the effective potential of low-energy
model (including one-loop and thermal corrections) and spectrum of its
scalar sector. In Section~\ref{numres} we present numerical results on
possibility of the first order EWPT and calculate spectrum of
gravitational waves generated during this transition. Section~\ref{conclusions}
is left for conclusions and discussion. 

%%%%%%%%%%%%%%%%%%%%%%%%%%%%%%%%%%%%%%%%%%%%%%%%%%%%%%%%%%%%%%%%%
\section{MSSM with goldstino supermultiplet} \label{SUSYmodel}
We consider a supersymmetric model in which the Lagrangian of Minimal
Supersymmetric Standard Model (MSSM) is extended by interactions involving
goldstino superfield $\Phi$. Since the auxiliary field $F_\Phi$ of
sgoldstino multiplet acquires non-zero vev $F$ upon SSB, the interaction
of $\Phi$ with MSSM fields yields soft supersymmetry breaking terms. The Lagrangian of the model can be written as
a sum~(see, e.g.~\cite{Gorbunov:2001pd})
\begin{equation}
  \label{eq:1}
\Lagr=\Lagr_{K}+\Lagr_{W}+\Lagr_{\text{gauge}}+\Lagr_{\Phi},
\end{equation}
where the terms involving K\"{a}hler potential $\Lagr_{K}$ read
\begin{equation}
\label{sfermion-masses}
\Lagr_{K}=\int d^2\theta d^2\bar{\theta}\sum_{k}\left(1- \frac{m_k^2}{F^2}\Phi^\dagger\Phi \right) \Phi_k^\dagger e^{g_1 V_1 + g_2 V_2 + g_3 V_3}\Phi_k, 
\end{equation}
here the sum is taken over all chiral matter superfields  $\Phi_k$, $m_k$ are soft
masses, $g_k$ and $V_k$ are the coupling constants and vector
superfields of the $U(1)_Y$, $SU(2)_W$, $SU(3)_c$ gauge groups. 
The part of the Lagrangian $\Lagr_{W}$ coming from superpotential of
the model has the form
\begin{multline}
\label{trilinear-terms}
\Lagr_{W}=\int d^2\theta\epsilon_{ij}\left( \left(\mu-\frac{B}{F}\Phi\right) H_d^i
H_u^j+\left(Y_{ab}^L+\frac{A^{L}_{ab}}{F}\Phi\right) L_a^j E_b^c H_d^i+\right.\\ \left.+\left(
Y_{ab}^D+\frac{A^{D}_{ab}}{F}\Phi\right) Q_a^j D_b^c H_d^i+\left(Y_{ab}^U+\frac{A^{U}_{ab}}{F}\Phi\right) Q_a^iU_b^cH_u^j
\right)+h.c.,  
\end{multline}
here $\mu$ is the higgsino mixing parameter, $L, E$ are the left
and right lepton superfields, $Q, U, D$ are superfields of left,
right up and down quarks, $H_u, H_d$ are the Higgs doublet 
superfields, $Y^{L, D, U}_{ab}$ are the Yukawa matrices, $A^L_{ab}$,
$A^D_{ab}$, $A^U_{ab}$ are soft trilinear constants; $a,b=1,2,3$ and superscript 'c' refers to the charged conjugated quantities.  
The Lagrangian involving gauge vector superfields $\Lagr_{\text{gauge}}$ is
\begin{equation}
\label{gaugino-masses}
\Lagr_{gauge}=\frac{1}{4}\sum_{a}\int d^2\theta\left(1+\frac{2M_a}{F}\Phi \right) \textnormal{Tr}W_{a\,\alpha}W^{\alpha}_a + h.c.\,, 
\end{equation}
where the sum goes over all the gauge groups of SM ($a=1,2,3$ for $U(1)_Y$, $SU(2)_W$, $SU(3)_c$), while
$M_1$, $M_2$, $M_3$ stand for the corresponding gaugino masses.   

When $\Phi$ \eqref{chiral} with $\langle F_\Phi\rangle=F$ is substituted into eqs.\,\eqref{sfermion-masses}--\eqref{gaugino-masses}, the terms proportional to $\Phi$ (and their Hermitian conjugates) in \eqref{trilinear-terms} and \eqref{gaugino-masses} provide soft trilinear terms and gaugino masses, while terms in \eqref{sfermion-masses} proportional to $\Phi^\dagger\Phi$ yield soft masses squared for the model scalars, i.e. Higgs bosons and superpartners of the SM fermions. These terms also induce goldstino couplings to the SM particles and their superpartners; we ignore the latter in our study, concentrating on the models where the superpartners are very heavy. Finally, these terms give rise to sgoldstino couplings to the SM particles which we include in our study.     

Then we introduce the Lagrangian of sgoldstino multiplet 
\begin{multline}\label{eq:lagr_phi}
\Lagr_\Phi=\int d^2\theta d^2\bar{\theta} \left(
\Phi^{\dagger}\Phi-\frac{\widetilde{m_s}^2+\widetilde{m_p}^2}{8F^2}
(\Phi^{\dagger}\Phi)^2-\frac{\widetilde{m_s}^2-\widetilde{m_p}^2}{12F^2}(\Phi^{\dagger}\Phi^3
+ \Phi^{\dagger 3}\Phi) -\right.\\\left.-
\frac{\delta_{\lambda_2}}{4F^2}H_u^{\dagger}H_u (\Phi^{\dagger}\Phi)^2
- \frac{\delta_{\lambda_3}}{9F^2}(\Phi^{\dagger}\Phi)^3 -
\frac{\delta_{\lambda_4}}{3F^2} H_u^{\dagger}H_u (\Phi^{\dagger}\Phi^3 
	+ \Phi^{\dagger 3}\Phi)-\right.\\\left.-
        \frac{\delta_{\lambda_5}}{5F^2}(\Phi^{\dagger}\Phi^5+\Phi^{\dagger
          5}\Phi) - \frac{\delta_{\lambda_6}}{8F^2}(\Phi^{\dagger
          2}\Phi^4+\Phi^{\dagger 4}\Phi^2)- 
        \frac{\delta_{\mu_1}}{2F^2}H_u^{\dagger}H_u
        (\Phi^{\dagger}\Phi^2 + \Phi^{\dagger 2}\Phi) -\right.\\\left.
        - \frac{\delta_{\mu_2}}{6F^2}(\Phi^{\dagger
          3}\Phi^2+\Phi^{\dagger 2}\Phi^3) -
        \frac{\delta_{\mu_3}}{4F^2}(\Phi^{\dagger
          4}\Phi+\Phi^{\dagger}\Phi^4) -
        \frac{\delta_{C_3}}{2F^2}(\Phi^{\dagger
          2}\Phi+\Phi^{\dagger}\Phi^2)\right)-\\-\left( \int d^2\theta
        F\Phi + h.c. \right).   
\end{multline}
Here $\widetilde{m_s}^2$ and $\widetilde{m_p}^2$ are the mass
parameters of scalar and pseudoscalar sgoldstinos, which become explicit upon substituting \eqref{chiral} with $\langle F_\Phi\rangle=F$. 

We note in passing that the MSSM soft parameters as well as other new model parameters entering \eqref{eq:lagr_phi} 
can be generically  complex-valued. In the present study we do
not discuss CP-violation and assume all the relevant parameters to be
real. Also, the described setup should be considered as a low-energy effective
theory valid at energies below $\sqrt{F}$. It is in
the perturbative regime as far as all the soft supersymmetry breaking terms are smaller than $\sqrt{F}$.

Most studies of the goldstino supermultiplet phenomenology limit themselves to the first three terms in \eqref{eq:lagr_phi}   
and the very last one triggering SSB in the model. Here we add new set of higher dimensional
operators with couplings constants of the same 
order $\frac{1}{F^2}$ as the second and third terms in~\eqref{eq:lagr_phi}. Therefore, they may emerge in the same way, e.g. technically upon integrating out all the heavy fields in the hidden sector. Parameters $\delta_{\lambda_i}$, $i=2,\dots,6$ are dimensionless, 
parameters $\delta_{\mu_i}$, $i=1,2$ have dimension of mass, while $\delta_{C_3}$ has dimension of squared mass. We will discuss their role in due course.

%%%%%%%%%%%%%%%%%%%%%%%%%%%%%%%%%%%%%%%%%%%%%%%%%%%%%%%%%%%%%%%
\section{Effective potential of the model}
\label{Sec:potential}
In what follows we are going to study the electroweak phase transition 
in the early Universe within this model. Properties of the EWPT are determined by the effective potential 
at the non-zero finite temperature. This potential can be calculated by making use of 
different approaches. In our study we use the perturbation theory and calculate the 
one-loop effective potential including finite temperature
corrections. 

The tree-level scalar potential of the model is 
\begin{equation} \label{eq:V}
V=V_{11}+V_{12}+V_{21}+V_{22},
\end{equation}
\begin{equation} \label{eq:V11}
V_{11}=\frac{g_1^2}{8}
\left(1+\frac{M_1}{F}(\phi+\phi^*)\right)^{-1}\left[h_d^\dagger
  h_d-h_u^\dagger h_u-\frac{\phi^*\phi}{F^2}\left(m_d^2h_d^\dagger
  h_d-m_u^2h_u^\dagger h_u\right)\right]^2, 
\end{equation}
\begin{equation} \label{eq:V12}
V_{12}=\frac{g_2^2}{8}
\left(1+\frac{M_2}{F}(\phi+\phi^*)\right)^{-1}\left[h_d^\dagger
  \sigma_a h_d+h_u^\dagger \sigma_a
  h_u-\frac{\phi^*\phi}{F^2}\left(m_d^2h_d^\dagger \sigma_a
  h_d+m_u^2h_u^\dagger \sigma_a h_u\right)\right]^2, 
\end{equation}
\begin{multline} 
V_{21}=\left(1-\frac{\widetilde{m_s}^2+\widetilde{m_p}^2}{2F^2}\phi^*\phi-\frac{\widetilde{m_s}^2-\widetilde{m_p}^2}{4F^2}(\phi^2+\phi^{*2})-\frac{m_u^2}{F^2}h_u^\dagger
h_u-\frac{m_d^2}{F^2}h_d^\dagger h_d-\right.\\\left.-\frac{m_u^4}{F^4}
h_u^\dagger h_u \phi^*\phi-\frac{m_d^4}{F^4} h_d^\dagger h_d
\phi^*\phi - \frac{\delta_{\lambda_2}}{F^2}h^\dagger_u h_u \phi^*\phi
- \frac{\delta_{\lambda_3}}{F^2}(\phi^*\phi)^2 -
\frac{\delta_{\lambda_4}}{F^2} h^\dagger_u h_u
(\phi^2+\phi^{*2})-\right.\\\left.-\frac{\delta_{\lambda_5}}{F^2}(\phi^4+\phi^{*4})-\frac{\delta_{\lambda_6}}{F^2}\phi^*\phi(\phi^2+\phi^{*2})-\frac{\delta_{\mu_1}}{F^2}
h^\dagger_u h_u (\phi+\phi^*)
-\frac{\delta_{\mu_2}}{F^2}\phi^*\phi(\phi+\phi^{*}) -
\right.\\\left. - \frac{\delta_{\mu_3}}{F^2}(\phi^3+\phi^{*3}) -
\frac{\delta_{C_3}}{F^2}(\phi+\phi^*)
\right)^{-1}\times\left|F+\left(-h_d^0
h_u^0+H^-H^+\right)\times\right.\\\left.\times\left(\frac{B}{F}-\frac{m_u^2+m_d^2}{F^2}\phi^*\left(\mu-\frac{B}{F}\phi\right)-\frac{\delta_{\mu_1}}{2F^2}\phi^*(2\phi+\phi^*)\mu\right)\right|^2, 
\end{multline}
\begin{equation} \label{eq:V22}
V_{22}=\frac{\mu^2\phi\phi^*}{F^2}\left(m_u^2h_d^\dagger h_d+m_d^2h_u^\dagger h_u\right)+\abs{\mu-\frac{B}{F}\phi}^2 \left(h_d^\dagger h_d+h_u^\dagger h_u\right).
\end{equation}
Here $h_d=\pmqty{h_d^0 \\ H^-}$, $h_u=\pmqty{H^+ \\ h_u^0}$ are the
Higgs doublets. 
The above expressions agree with the potential found
in~\cite{Demidov:2020jne} if one puts all the '$\delta$'-couplings entering~\eqref{eq:lagr_phi} to zero. Note that
in the Lagrangian each power of sgoldstino field comes with the power
of $1/F$, therefore for $\sqrt{F}$ to be considerably larger than sgoldstino masses and soft SUSY breaking parameters the higher order interaction terms are parametrically suppressed; consequently, in \eqref{eq:V}--\eqref{eq:V22} we neglect all the terms suppressed by $1/F^3$ 
and stronger.  For the same reason, in this work we limit ourselves to interactions of the Higgs doublets to sgoldstino up to the second power in the latter.

In this study we consider a scenario where all superpartners
of the SM particles are considerably heavier than sgoldstinos, which are assumed to
have masses of the order of the electroweak scale. For the Higgs sector it
corresponds to the decoupling regime of MSSM.
At low energies we are left with a single Higgs doublet $\mathcal{H}$
instead of $h_u$ and $h_d$: all the MSSM Higgs bosons are heavy except for the SM-like Higgs field.  The low-energy effective Lagrangian can
be obtained via the following substitution
\begin{equation}
  \label{eq:decoupling}
h_u\to \mathcal{H}\sin{\beta}, \;\;
h_d\to -\epsilon \mathcal{H}^*\cos{\beta}\,.
\end{equation}
Here $\tan{\beta}=v_u/v_d$ with $v_u$ and $v_d$ being vacuum
expectation values of neutral components of the Higgs doublets,
$h_u$ and $h_d$, respectively.
The tree-level potential at zero temperature can be written in terms
of the Higgs doublet $\mathcal{H}$ and the complex scalar sgoldstino
field $\phi$ as 
\begin{equation} \label{eq:Vtree}
V_{\text{tree}}(\mathcal{H}, \phi) = V_{\text{quartic}}(\mathcal{H}, \phi)+V_{\text{cubic}}(\mathcal{H}, \phi)+V_{\text{free}}(\mathcal{H}, \phi)\,,
\end{equation}
where 
\begin{multline} \label{eq:Vquartic}
V_{\text{quartic}}(\mathcal{H}, \phi)= \lambda_1 (\mathcal{H}^\dagger \mathcal{H})^2+\lambda_2\phi^*\phi \mathcal{H}^\dagger \mathcal{H} + \lambda_3 (\phi^*\phi)^2+\lambda_4(\phi^2+\phi^{*2})\mathcal{H}^\dagger \mathcal{H}+\\+\lambda_5(\phi^4+\phi^{*4})+\lambda_6\phi^*\phi(\phi^2+\phi^{*2})\,,
\end{multline}
\begin{equation} \label{eq:Vcubic}
V_{\text{cubic}}(\mathcal{H}, \phi)= \frac{\mu_1}{\sqrt{2}}(\phi+\phi^*)\mathcal{H}^\dagger \mathcal{H} + \frac{\mu_2}{\sqrt{2}}(\phi+\phi^*)\phi^*\phi + \frac{\mu_3}{\sqrt{2}}(\phi^3+\phi^{*3})\,,
\end{equation}
\begin{equation} \label{eq:Vfree}
V_{\text{free}}(\mathcal{H}, \phi)= -\widetilde{M_1}^2\mathcal{H}^\dagger \mathcal{H} + \widetilde{M}_2^2\phi^*\phi + \widetilde{M}_3^2\left(\phi^2 + \phi^{*2} \right)+ \frac{C_3}{\sqrt{2}}(\phi+\phi^*)\,.
\end{equation}
The coupling constants which appear above are related to the
parameters of the model Lagrangian as follows  
\begin{multline}
  \label{eq:coupl1}
\lambda_2 = \frac{\delta_{\lambda_2}}{2}(1-\cos 2\beta)+\frac{\delta_{\mu_1}\mu}{F}\sin 2\beta+\frac{1}{F^2}\left[2\delta_{C_3}\delta_{\mu_1}(1-\cos 2\beta)+\right.\\\left.+\frac{\mu^2 m_A^2}{2}(1+\cos^2 2\beta)+\frac{m_Z^4}{4}\cos^2 2\beta+(\widetilde{m_s}^2+\widetilde{m_p}^2)\left(\frac{m_A^2}{4}\sin^2 2\beta -\mu^2\right)+\right.\\\left.+ \frac{m_Z^2}{2} \cos^2 2\beta \left(3\mu^2 - \widetilde{m_s}^2-\widetilde{m_p}^2\right)\right]\,, 
\end{multline}
\begin{equation}
  \label{eq:coupl2}
\lambda_3 = \delta_{\lambda_3} + \frac{1}{F^2}\left[\frac{1}{4}(\widetilde{m_s}^2+\widetilde{m_p}^2)^2 + \frac{1}{8}(\widetilde{m_s}^2-\widetilde{m_p}^2)^2 + 4\delta_{\mu_2}\delta_{C_3} \right], 
\end{equation}
\begin{multline} 
  \label{eq:coupl3}
\lambda_4= \frac{\delta_{\lambda_4}}{2}(1-\cos 2\beta) + \frac{\delta_{\mu_1}\mu}{4F}\sin 2\beta+\frac{\delta_{C_3}\delta_{\mu_1}}{F^2}(1-\cos 2\beta)+\\+\frac{\widetilde{m_s}^2-\widetilde{m_p}^2}{2F^2}\left[\frac{m_A^2}{4}\sin^2 2\beta-\mu^2 - \frac{m_Z^2}{2} \cos^2 2\beta\right]\,, 
\end{multline}
\begin{equation}
  \label{eq:coupl4}
\lambda_5 = \delta_{\lambda_5} + \frac{1}{F^2}\left[\frac{(\widetilde{m_s}^2-\widetilde{m_p}^2)^2}{16}+2\delta_{\mu_3}\delta_{C_3} \right]\,, 
\end{equation}
\begin{equation}
  \label{eq:coupl5}
\lambda_6 = \delta_{\lambda_6} + \frac{1}{F^2}\left[\frac{\widetilde{m_s}^4-\widetilde{m_p}^4}{4} + 2(\delta_{\mu_2}+\delta_{\mu_3})\delta_{C_3} \right]\,, 
\end{equation}
\begin{equation} 
  \label{eq:coupl6}
\frac{\mu_1}{\sqrt{2}} = \frac{\delta_{\mu_1}}{2}(1-\cos 2\beta)-\frac{\mu^3}{F}\sin 2\beta+\frac{\delta_{C_3}}{F^2}\left[\frac{m_A^2}{2}\sin^2 2\beta-2\mu^2-m_Z^2\cos^2 2\beta \right]\,,
\end{equation}
\begin{equation}
  \label{eq:coupl7}
\frac{\mu_2}{\sqrt{2}} = \delta_{\mu_2} + \frac{\delta_{C_3}}{2F^2}(3\widetilde{m_s}^2+\widetilde{m_p}^2)\,,
\end{equation}
\begin{equation}
  \label{eq:coupl8}
\frac{\mu_3}{\sqrt{2}} = \delta_{\mu_3} + \frac{\delta_{C_3}}{2F^2}(\widetilde{m_s}^2-\widetilde{m_p}^2)\,,
\end{equation}
\begin{equation} 
  \label{eq:coupl9}
\widetilde{M}_2^2=\frac{\widetilde{m_s}^2+\widetilde{m_p}^2}{2}+\frac{2\delta^2_{C_3}}{F^2}\,,
\end{equation}
\begin{equation} 
  \label{eq:coupl10}
\widetilde{M}_3^2=\frac{\widetilde{m_s}^2-\widetilde{m_p}^2}{4}+\frac{\delta^2_{C_3}}{F^2}\,,
\end{equation}
\begin{equation}
  \label{eq:coupl11}
\frac{C_3}{\sqrt{2}}=\delta_{C_3}\,.
\end{equation} 
These expressions are obtained from\,\eqref{eq:V}  by substituting\, \eqref{eq:decoupling} and expanding in powers of $1/F$; all the terms suppressed by $1/F^3$
and stronger are neglected.

Next, we express the scalar fields as follows
\begin{equation} \label{eq:doublet}
\mathcal{H} = \frac{1}{\sqrt{2}}
\begin{pmatrix}
G^+ \\ h + iG^0
\end{pmatrix},\;\;\;
\phi=\frac{1}{\sqrt{2}}(s+ip)\,.
\end{equation}
Here $h$ is the Higgs field while real field $G^0$ and complex field
$G^+$ are Goldstone bosons, $s$ and $p$ are scalar and pseudoscalar
sgoldstinos. Substituting formulas \eqref{eq:doublet} into
\eqref{eq:Vtree}--\eqref{eq:Vfree} we obtain the
tree-level zero-temperature potential $V_0(h, s, p)$ of three scalar
fields $h$, $s$, $p$
\begin{multline} \label{eq:V0}
V_0(h, s, p)=\frac{\lambda_1}{4} h^4 +\frac{\lambda_{hs}}{4}h^2 s^2 +\frac{\lambda_{hp}}{4}h^2 p^2+\frac{\lambda_s}{4} s^4 +\frac{\lambda_p}{4} p^4 +\frac{\lambda_{sp}}{4}s^2p^2+\\
+ \frac{\mu_1}{2}sh^2 + \frac{\mu_s}{6}s^3 + \frac{\mu_{sp}}{2}sp^2 - \frac{\widetilde{M}_1^2}{2} h^2 +\frac{M_s^2}{2} s^2+\frac{M_p^2}{2} p^2 + C_3s\,.
\end{multline} 
Here for convenience we introduce new coupling constants  
\begin{equation}
\begin{matrix}
\lambda_{hs}\equiv\lambda_2+2\lambda_4\,, && \lambda_s\equiv\lambda_3+2\lambda_5+2\lambda_6\,,  && \lambda_p\equiv\lambda_3+2\lambda_5-2\lambda_6\,,\\
\lambda_{hp}\equiv\lambda_2-2\lambda_4\,, && \mu_s\equiv 3(\mu_2+\mu_3)\,, && \mu_{sp}\equiv \mu_2-3\mu_3\,, \\
\lambda_{sp}\equiv 2\lambda_3-12\lambda_5, && M_s^2 \equiv \widetilde{M}_2^2 + 2\widetilde{M}_3^2\,, && M_p^2 \equiv \widetilde{M}_2^2 - 2\widetilde{M}_3^2\,.
\end{matrix}
\end{equation}
Let us note that in~\eqref{eq:V0} the terms with Goldstone bosons
have been omitted. We will take them into account later on in the
calculation of field-dependent particle masses and the effective
potential. 

In the present scenario we assume that sgoldstino field does not acquire
non-zero vacuum expectation value at zero temperature. In order to fix
position of the minimum of tree level potential $V_0(h, s, p)$ at
the point $\left<h\right>=v$, $\left<s\right>=\left<p\right>=0$,
with $v=246 \text{ GeV}$, in what follows we set parameters $\widetilde{M}_1^2$
and $C_3$ to be   
\begin{equation}
  \label{eq:fix}
\widetilde{M}_1^2=\lambda_1 v^2, \quad\quad C_3=-\mu_1 v^2/2\,.
\end{equation} 

One-loop corrections to the potential $V_0(h, s, p)$ at zero temperature
in the form of Coleman-Weinberg potential in $\overline{MS}$-scheme
\cite{Coleman:1973jx} look as  
\begin{equation}
  \label{eq:CW}
V_{CW}(h, s, p)=\frac{1}{64\pi^2} \sum_i (-1)^{s_i} n_i m_i^4(h, s, p)
\left(\log{\frac{m_i^2(h, s, p)}{Q^2}}-c_i \right),  
\end{equation}
where the sum goes over all fields of the model; $Q$ is the renormalization 
scale and in our calculations we set $Q=100\text{ GeV}$. Contributions from
bosons come with ``$+$" sign, i.e. $s_i=0$, fermion terms are summed with
``$-$" sign, $s_i=1$. In~\eqref{eq:CW} the particle masses $m_i(h, s, p)$ depend
on background values of the scalar fields $h$, $s$, $p$. For each particle
$n_i$ is the number of its degrees of freedom, $c_i=3/2$ for scalar particles
and fermions, $c_i=5/6$ for massive vector particles.

Field-dependent squared masses of the scalars $h$, $s$, $p$ are the eigenvalues of the Hessian matrix of the potential \eqref{eq:V0},  
\begin{equation}\label{mmatrix}
\begin{pmatrix}
V_{0,hh} & V_{0,hs} & V_{0,hp} \\
V_{0,hs} & V_{0,ss} & V_{0,sp} \\
V_{0,hp} & V_{0,sp} & V_{0,pp} \\
\end{pmatrix},
\end{equation}
where 
\begin{equation}
V_{0,hh}=3\lambda_1h^2+\frac{\lambda_{hs}}{2}s^2+\mu_1s+\frac{\lambda_{hp}}{2}p^2-\widetilde{M}_1^2\,,
\end{equation}
\begin{equation}
V_{0,hs}=\lambda_{hs}hs+\mu_1h\,,
\end{equation}
\begin{equation}
V_{0,hp}=\lambda_{hp}hp\,,
\end{equation}
\begin{equation}
V_{0,ss}=\frac{\lambda_{hs}}{2}h^2+3\lambda_s s^2+\frac{\lambda_{sp}}{2} p^2+\mu_s s+M_s^2\,,
\end{equation}
\begin{equation}
V_{0,sp}=\lambda_{sp}sp+\mu_{sp}p\,,
\end{equation}
\begin{equation}
V_{0,pp}=\frac{\lambda_{hp}}{2} h^2 + \frac{\lambda_{sp}}{2} s^2 + 3\lambda_p p^2 + \mu_{sp}s+M_p^2\,.
\end{equation}
In particular, we obtain the following expressions for the Higgs boson,
sgoldstino scalar and pseudoscalar squared masses at zero temperature
near the electroweak vacuum $(v, 0, 0)$:    
\begin{equation} \label{eq:mhs}
m^{2}_{h,s~phys} = \lambda_1 v^2 +\frac{\lambda_{hs}}{4}v^2 +
\frac{M_s^2}{2} \pm \sqrt{\left(\lambda_1 v^2
  -\frac{\lambda_{hs}}{4}v^2 - \frac{M_s^2}{2} \right)^2 + \mu_1^2
  v^2}\,, 
\end{equation}
\begin{equation} \label{eq:mp}
m^2_{p~phys} = \frac{\lambda_{hp}}{2}v^2 + M_p^2\,.
\end{equation}
The tree-level squared field-dependent masses of top-quark, $Z$,
$W^{\pm}$-bosons are as usual 
\begin{equation}
m_t^2(h)=\frac{y_t^2}{2}h^2\,, \quad m_Z^2(h)=\frac{g^{'\,2}+g^2}{4}h^2\,,
\quad m_W^2(h)=\frac{g^2}{4}h^2\,, 
\end{equation}
here $y_t$ is the Yukawa coupling constant of top-quark, $g'$ and
$g$ are the $U(1)_Y$ and $SU(2)_W$ coupling constants, while squared
mass of Goldstone bosons $G^0$, $G^\pm$ is 
\begin{equation}
m^2_G(h, s, p) =
\lambda_1h^2+\frac{\lambda_{hs}}{2}s^2+\mu_1s+\frac{\lambda_{hp}}{2}p^2- \widetilde{M}_1^2\,. 
\end{equation}

In general, quantum corrections to the effective potential represented 
by $V_{CW}(h, s, p)$ shift the position of the tree level
potential minimum. In order to fix the minimum at zero
temperature at the point $(v, 0, 0)$ as well as to fix values of
scalar particle masses in this minimum, we add the following
counterterms  
\begin{equation}
V_{CT}(h, s, p) = \frac{\delta\lambda_1}{4} h^4
+\frac{\delta\lambda_{hs}}{4}h^2 s^2 +\frac{\delta\lambda_{hp}}{4}h^2
p^2+ \frac{\delta\mu_1}{2}sh^2 - \frac{\delta \widetilde{M}_1^2}{2} h^2 + \delta
C_3s\,. 
\end{equation}
Coefficients $\delta\lambda_1, \delta\lambda_{hs}, \delta\lambda_{hp},
\delta\mu_1, \delta \widetilde{M}_1^2, \delta C_3$ in the counterterm potential $V_{ct}(h,s,p)$ are chosen in such a way that all the first and the second partial derivatives of the sum $V_{CW}+V_{ct}$ with respect to fields $h$, $s$, $p$ are equal to zero at the point $(v, 0, 0)$ \cite{Chiang:2019oms}, which results in 
\begin{equation}
\delta \lambda_1= -\frac{1}{2v^2}\eval{\pdv[2]{V_{CW}}{h}}_{(v, 0, 0)} + \frac{1}{2v^3}\eval{\pdv{V_{CW}}{h}}_{(v, 0, 0)}\,, 
\end{equation}
\begin{equation}
\delta\lambda_{hs}=-\frac{2}{v^2}\eval{\pdv[2]{V_{CW}}{s}}_{(v, 0, 0)}, \quad\quad \delta\lambda_{hp}=-\frac{2}{v^2}\eval{\pdv[2]{V_{CW}}{p}}_{(v, 0, 0)}\,,
\end{equation} 
\begin{equation}
\delta\mu_1=-\frac{1}{v}\eval{\pdv{V_{CW}}{h}{s}}_{(v, 0, 0)}\,, \quad
\delta C_3 = -\eval{\pdv{V_{CW}}{s}}_{(v, 0, 0)}+\frac{v}{2}\eval{\pdv{V_{CW}}{h}{s}}_{(v, 0, 0)}\,,
\end{equation}
\begin{equation}
\delta \widetilde{M}_1^2=-\frac{1}{2}\eval{\pdv[2]{V_{CW}}{h}}_{(v, 0, 0)} + \frac{3}{2v}\eval{\pdv{V_{CW}}{h}}_{(v, 0, 0)}\,.
\end{equation}
Full expression for the scalar potential of the model at zero temperature reads 
\begin{equation} \label{eq:VzeroT}
V_{T=0}(h, s, p) = V_0(h, s, p) + V_{CW}(h, s, p) + V_{CT}(h, s, p).
\end{equation}

To summarize, with all the assumptions described above the model scalar
potential\,\eqref{eq:V0} is fully specified by the
following parameters  
\begin{equation}
\lambda_s,\;\;
\lambda_p,\;\;
\lambda_{hs},\;\;
\lambda_{hp},\;\;
\lambda_{sp},\;\;
\lambda_{hp},\;\;
\mu_1,\;\;
\mu_s,\;\;
\mu_{sp}\,,
\end{equation}
as well as masses $m_{s\,phys}$,
$m_{p\,phys}$ and $m_{h\,phys}=125$\,GeV. As an input for the analysis of the EWPT dynamics we
use these parameters taken directly at the electroweak scale, i.e. at
$Q\sim 100$~GeV, while other quantities entering~\eqref{eq:V0} can be
found from~\eqref{eq:fix} which fixes the position of the minimum at zero
temperature and the expressions~\eqref{eq:mhs} and~\eqref{eq:mp}. In due course 
we will discuss relation to the parameters of supersymmetric model. 

The one-loop effective potential at finite temperature includes
the following thermal corrections (the free energy density of plasma)
\cite{Dolan:1973qd} 
\begin{equation} \label{eq:VT}
V_T(T, h, s, p)=\frac{T^4}{2\pi^2} \sum_i n_i
J_{B/F}\left(\frac{m_i(h, s, p)}{T} \right)\,,  
\end{equation}
where $T$ is temperature, and the sum is taken over all particle species
that are in thermal equilibrium in plasma. Special thermal functions
$J_B(x)$ for bosons and $J_F(x)$ for fermions are defined as   
\begin{equation} \label{eq:funcJ}
J_{B/F}(x)=\pm \int_0^{\infty} \dd{y} y^2 \ln{\left(1\mp \exp(-\sqrt{x^2+y^2})\right)}\,.
\end{equation}
The integral \eqref{eq:funcJ} cannot be expressed in terms of
analytical functions, although it can be expanded into series of
MacDonald functions 
\begin{equation}
J_{B/F}(x)=\mp x^2 \sum_{n=1}^\infty \frac{(\pm 1)^n}{n^2} K_2(nx)\,.
\end{equation}

It is known that validity of the perturbation theory for the effective
potential breaks down at high temperatures. This problem can be cured
by resummation of daisy diagrams (see,
e.g.~\cite{Carrington:1991hz,Arnold:1992rz}) 
which yields the following additional contribution to the effective
potential 
\begin{equation} \label{eq:Vd}
    V_d(T, h, s, p)=-\frac{T}{12\pi} \sum_i a_i n_i \left(\left( m_{T\,i}^2(T,h,s,p)\right)^{3/2}-\left( m_i^2(h,s,p)\right)^{3/2} \right),
\end{equation}
where $a_i=1$ for scalars and $a_i=1/3$ for vector bosons. While
ordinary particle masses $m_i^2$ are eigenvalues of the mass matrix
$\mathcal{M}^2$, the thermal Debye masses $m_{T\,i}^2$ are the eigenvalues of the full mass matrix $\mathcal{M}^2+\Pi(T)$ that includes thermal
corrections. For the Higgs boson $h$ and sgoldstinos $s$, $p$ the thermal
corrections to the squared mass matrix have the form $\Pi(T)=\text{diag}(\Pi_h, \Pi_s,
\Pi_p)$, where   
\begin{equation} \label{eq:Ph}
    \Pi_h(T)=T^2 \left(
    \frac{\lambda_1}{2}+\frac{\lambda_{hs}+\lambda_{hp}}{24}+\frac{3g^2+g^{'\,2}}{16}+\frac{y_t^2}{4}\right), 
\end{equation}
\begin{equation} \label{eq:Ps}
    \Pi_s(T)=T^2 \left(\frac{\lambda_{hs}}{6} + \frac{\lambda_s}{4} + \lambda_{sp}\right),
\end{equation}
\begin{equation} \label{eq:Pp}
    \Pi_p(T)=T^2 \left(\frac{\lambda_{hp}}{6} + \frac{\lambda_p}{4} + \lambda_{sp}\right).
\end{equation}
Debye corrections for the Goldstone bosons are 
\begin{equation}
    \Pi_{G^0, G^{\pm}}(T)= T^2 \left(\frac{\lambda_1}{2} + \frac{\lambda_{hs}+\lambda_{hp}}{24} + \frac{3g^2+g^{'\,2}}{16}+\frac{y_t^2}{4}\right).
\end{equation}
Masses of transverse components of the vector bosons do not acquire Debye
corrections, while the mass squared of longitudinal component of
$W$-boson receive thermal correction $\Pi_{W_{L}}(T)= \frac{11}{6} g^2 T^2$. 
The full mass matrix of photon and $Z$-boson longitudinal components
is 
\begin{equation}
    \mathcal{M}^2_{\gamma Z_{L}}+\Pi_{\gamma Z_{L}}(T) = 
    \begin{pmatrix}
    \frac{1}{4} g^2 h^2+\frac{11}{6}g^2T^2 & \frac{1}{4} gg' h^2 \\
    \frac{1}{4} gg' h^2 & \frac{1}{4} g^{'\,2} h^2+\frac{11}{6}g^{'\,2}T^2
    \end{pmatrix}
\end{equation}
Sum of \eqref{eq:VzeroT}, \eqref{eq:VT} and \eqref{eq:Vd} gives the
one-loop finite-temperature effective potential of the model 
\begin{equation} \label{eq:Vtot}
V_{\text{eff}}(T, h, s, p)=V_{T=0}(h, s, p)+V_T(T, h, s, p)+V_d(T, h,
s, p). 
\end{equation}

Let us comment briefly on known theoretical uncertainties inherent 
in utilizing the effective potential in the form~\eqref{eq:Vtot}. Firstly, 
it is known that the effective potential is explicitly gauge dependent~\cite{Patel:2011th,Garny:2012cg} and we use the effective potential calculated in the Landau gauge. 
Secondly, we truncate the perturbative expansion, keeping only  the one-loop order terms, which imposes the dependence on the 
renormalization scale. The two-loop analysis of EWPT in a
singlet-extended SM performed recently~\cite{Niemi:2021qvp}
has revealed that the two-loop corrections may change the values of the critical temperature and the latent heat by 20--50\% (up to 100\%) depending on the model. In respect to these issues, 
we refer to the recent review~\cite{Croon:2020cgk} for an extensive
discussion of various theoretical uncertainties typical in investigations of the cosmological first order
phase transitions. 
One should bear in mind those uncertainties when applying the numerical results presented in the rest of the paper.

%%%%%%%%%%%%%%%%%%%%%%%%%%%%%%%%%%%%%%%%%%%%%%%%%%%%%%%
\section{Numerical results} \label{numres}

In this Section we investigate the possibility of getting the
first order electroweak phase transition in the supersymmetric model
with relatively light sgoldstinos and calculate the spectrum of gravitational waves produced during this transition. 
Since the model parameter space is multidimensional and large, 
we do not perform scanning over it to indicate the regions where the EWPT is of the first order, but rather present several points in this space, where we checked the EWPT is of the first order and studied the generation of gravitational waves. 

As to the found benchmark points, we recall that generically the first order EWPT occurs in very specific parts of the model parameter
space and even small shifts of the parameters can change properties of the phase transition drastically. 
Nevertheless, the chosen points are in a rather compact region of the model parameter space. And as we found the first order EWPT in each  of them (though with different characteristics) we certainly can claim that this phenomenon is not unique for these sets of parameters. At the same time we calculate the spectra of gravitational waves produced during the EWPT in each case and found that even within this compact part of the parameter space the
model predicts wide range of peak frequency and amplitude of this
spectrum. Finally, we illustrated the impact of some theoretical uncertainties discussed in the previous Section on the model predictions.  

%%%%%%%%%%%%%%%%%%%%%%%%%%%%%%%%%%%%%%%%%%%%%%%%%%%%%%%%%
\subsection{First-order electroweak phase transition}
The dynamics of EWPT is determined by the effective
potential~\eqref{eq:Vtot}. The numerical package
\texttt{PhaseTracer}~\cite{Athron:2020sbe} was 
used to search for the models in the parameter space for which
the first-order electroweak phase transition takes place and explore  its properties. Besides, the thermal
functions~\eqref{eq:funcJ} were computed using the numerical library 
\texttt{thermal\_funcs}~\cite{Fowlie:2018eiu}. With the help of this
software we can obtain the value of potential~\eqref{eq:Vtot} and its
minima positions in the field space as functions of temperature
$T$. We select several benchmark models dubbed $Q_1,.., Q_{10}$ in which (as we have found) 
the first order phase transition takes place. 
In Table\,\ref{table:GWparams}
\begin{table}[!htb]
  \centering
  \begin{tabular}{|c|c|c|c|c|c|c|c|c|c|c|}
    \hline
    BM & $\lambda_{hs}$ & $\lambda_{hp}$ & $\lambda_s$ & $\lambda_p$ & $\lambda_{sp}$ & $T_{\text{nuc}}$ & $T_{\text{perc}}$ & $\alpha\cdot 10^3$ & $\beta/H_c$ & $g_*$\\
    \hline
    $Q_1$ & 0.88 & 0.78 & 0.70 & 0.60 & 0.70 & 63.58 & 58.30 & 119 & 254 & 107.25\\
    \hline
    $Q_2$ & 0.85 & 0.75 & 0.70 & 0.60 & 0.70 & 81.51 & 78.22 & $37.4$ & 602 & 107.75\\
    \hline
    $Q_3$ & 0.80 & 0.70 & 0.70 & 0.60 & 0.70 & 99.60 & 97.64 & $15.2$ & 1470 & 107.75\\
    \hline
    $Q_4$ & 0.75 & 0.65 & 0.70 & 0.60 & 0.70 & 112.25 & 111.05 & $8.61$ & 2920 & 107.75\\
    \hline
    $Q_5$ & 0.70 & 0.65 & 0.70 & 0.60 & 0.70 & 122.06 & 121.36 & $5.48$ & 5760 & 107.75\\
    \hline
    $Q_6$ & 0.65 & 0.60 & 0.70 & 0.60 & 0.70 & 130.07 & 129.72 & $3.52$ & 12700 & 107.75\\
    \hline
    $Q_7$ & 0.60 & 0.50 & 0.55 & 0.45 & 0.60 & 133.45 & 133.13 & $2.88$ & 15700 & 107.75\\
    \hline
    $Q_8$ & 0.60 & 0.50 & 0.60 & 0.50 & 0.70 & 134.84 & 134.61 & $2.56$ & 19400 & 107.75\\
    \hline
    $Q_9$ & 0.60 & 0.50 & 0.65 & 0.55 & 0.70 & 135.85 & 135.66 & $2.26$ & 27800 & 107.75\\
    \hline
    $Q_{10}$ & 0.60 & 0.40 & 0.70 & 0.60 & 0.70 & 136.74 & 136.61 & $1.95$ & 38100 & 107.75\\
    \hline
  \end{tabular}  \caption{Dimensionless parameters of selected benchmark models
    (BM) with the first order electroweak phase transition as
    well as nucleation and percolation temperatures $T_{\text{nuc}}$, $T_{\text{perc}}$ (in GeV), strength
    $\alpha$, ratio $\beta/H_c$ and effective number of degrees of
    freedom $g_*$. \label{table:GWparams}}
\end{table}
 we present the values, which dimensionless parameters 
$\lambda_{hs}$, $\lambda_{hp}$, $\lambda_{s}$, $\lambda_p$ and
$\lambda_{sp}$ take in these models.
As for other parameters, we fix $\mu_1=5$\,GeV, $\mu_s=30$\,GeV, $\mu_{sp}=-10$\,GeV and choose $m_{s~phys}=110$\,GeV,  $m_{p~phys}=440$\,GeV for concreteness. 
  
\begin{figure}[]
\begin{center}
  \includegraphics[width=\textwidth]{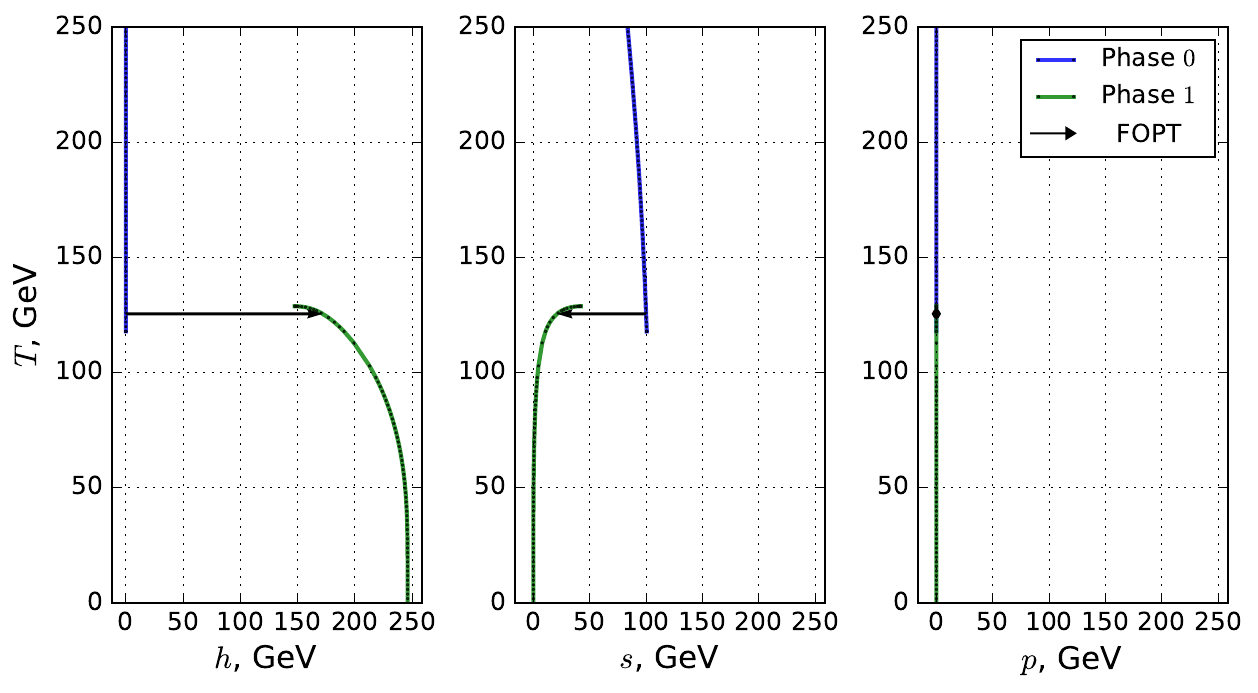}
  \caption{\label{fig:phi-T} Temperature dependence of field values
    at the minimum of the effective potential for the benchmark model
    $Q_5$, see Table~\ref{table:GWparams}. The full list of
    parameter values is given in the text. This graph is plotted
    using the package \texttt{PhaseTracer}~\cite{Athron:2020sbe}.}  
\end{center}
\end{figure}

  Let us note that
the first order phase transition in a SM extension with scalar singlet
typically requires large coupling between the Higgs boson and the
scalar, see 
e.g.~\cite{Kurup:2017dzf,Gould:2019qek,Chiang:2019oms,Cline:2021iff}. 
Indeed, the deformation of the Higgs effective potential, 
needed to replace the expected in the SM crossover with the first order EWPT, naturally implies a sufficiently strong Higgs interaction with the new scalar(s). Then the models with light enough sgoldstinos would admit the Higgs boson decays $h\to ss$ or $h\to pp$ with unacceptably large partial widths. On the other hand, the scalar potential with heavy sgoldstinos would not lead to the first order EWPT, because their contribution to the thermal dynamics will be suppressed by the mass, and the form of the Higgs effective potential remains similar to that in the SM.  Figure\,\ref{fig:phi-T} 
shows an example of minimum positions depending on temperature for the
benchmark point $Q_5$ with model parameters shown in
Table~\ref{table:GWparams}. In this example the value of pseudoscalar
sgoldstino field $\left\langle p \right\rangle= 0$ at any temperature. At
high temperatures there exists only a single phase with $\left\langle  
h\right\rangle=0$ and $\left\langle s\right\rangle \neq0$. At
$T=129\text{ GeV}$ another phase with broken electroweak symmetry,
i.e. when $\left\langle h\right\rangle \neq0$ and
$\left\langle s\right\rangle =0$, naturally appears. The arrow in
Figure~\ref{fig:phi-T} marks the  critical temperature
$T_{c}=126\text{ GeV}$ at which these two phases have equal values of
the effective potential. The electroweak phase transition for other
benchmark models in Table~\ref{table:GWparams} 
exhibit the similar dynamics. Let us note that despite $\left\langle p \right\rangle= 0$ at any temperature, the effective potential still depends on the dimensionless parameters $\lambda_{hp}$, $\lambda_p$, $\lambda_{sp}$ as they contribute to thermal corrections \eqref{eq:Ph}--\eqref{eq:Pp}.

The phase transition starts when bubbles of new phase begin to
nucleate in the Universe. The bubble production rate per unit volume
is given by~\cite{Linde:1980tt,Linde:1981zj}   
\begin{equation}
\label{exponent}
P\sim {\cal A}(T)\exp{-\frac{S_3}{T}},
\end{equation}
where ${\cal A}(T)$ is a dynamical prefactor and $S_3$ is
$O(3)$-symmetric Euclidean action 
\begin{equation}
S_3=\int_0^\infty 4\pi \rho^2 \dd{\rho} \left(V_{\text{eff}}(T, h, s,
p)+\frac{1}{2}\left(\dv{h}{\rho}\right)^2
+ \frac{1}{2}\left(\dv{s}{\rho}\right)^2 +
\frac{1}{2}\left(\dv{p}{\rho}\right)^2 \right)\,,
\end{equation}
calculated on the bounce solution, with $\rho$ being radial
coordinate. The bounce is a spherically symmetric
solution of the classical field equations~\cite{Dine:1992wr}
\begin{equation}
  \label{eq:bounce}
\dv[2]{\varphi_i}{\rho}+\frac{2}{\rho}\dv{\varphi_i}{\rho}
=\pdv{V_{\text{eff}}}{\varphi_i}\,, 
\end{equation}
where $\varphi_i$ stands for $h$, $s$ or $p$. The
equation~\eqref{eq:bounce} is supplemented with the following boundary 
conditions: 1) $\frac{\dd{\varphi_i}(\rho)}{\dd{\rho}} =0$ at $\rho=0$; 
2) $\varphi_i(\rho)\to\varphi_i^{\rm false}$ at $\rho\rightarrow \infty$,
where $\varphi_i^{\rm false}$ are values of the fields in the false
minimum. The nucleation in the primordial plasma starts at temperature $T_{\text{nuc}}$, when the factor in exponent\,\eqref{exponent} drops to $\frac{S_3}{T}\big|_{T=T_{\text{nuc}}}\simeq  140$~\cite{Anderson:1991zb}. 

\begin{figure}[!htb]
  \begin{center}
    \includegraphics[width=0.8\textwidth]{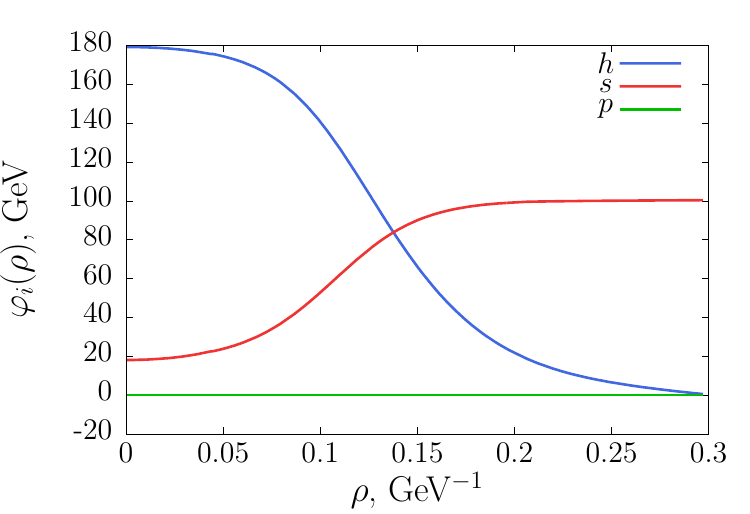}
    \caption{\label{fig:profile} Profile of the bounce solution for the
      benchmark model $Q_5$ at the nucleation temperature. Blue (top),
      red (middle) and green (bottom) curves correspond to $h$, $s$
      and $p$, respectively. The field $\varphi_i$ value in GeV is on
      the vertical axis, the distance to the origin in 3-dimensional
      space $\rho$ in GeV$^{-1}$ is on the horizontal axis. The origin
      is in the bubble center. This graph was plotted using the
      package~\texttt{FindBounce}~\cite{Guada:2020xnz}.} 
  \end{center}
\end{figure}

To find numerically the bounce solution and calculate the action $S_3$ entering~\eqref{exponent} we use the package \texttt{FindBounce}~\cite{Guada:2020xnz}. Figure\,\ref{fig:profile}  
shows an example of bounce profile in space at temperature
$T\approx 122\text{ GeV}$. To find the nucleation temperature
$T_{\text{nuc}}$ we solve numerically equation
$\frac{S_3}{T}\big|_{T=T_{\text{nuc}}}=140$. 

After the temperature reaches $T_{\text{nuc}}$, more and more bubbles
nucleate and the latent heat of the unbroken phase releases. The bubbles of new phase expand decreasing the fraction of the Universe in the metastable state. The phase transition is completed after the bubble percolation which happens at the temperature $T_{perc}$ to be found~\cite{Caprini:2019egz} from the equation
\begin{equation}
    \frac{S_3(T_{perc})}{T_{perc}} = 131 + \log{\left(\frac{A}{T^4}\right)} - 4\log{\left(\frac{T}{100~{\rm GeV}}\right)} - 4\log{\left(\frac{\beta(T_{perc})/H_c}{100}\right)} + 3\log{v_w}\,,
\end{equation}
where $\log{\left(\frac{A}{T^4}\right)}\approx -14$ for EWPT~\cite{Carrington:1993ng} and $v_w$ is the bubble wall velocity. 
In this study we take $v_w=0.55$ as an exemplary value assuming deflagration for the bubble growth regime (see e.g.~\cite{Moore:1995si,Konstandin:2014zta,Kozaczuk:2015owa,Friedlander:2020tnq} for discussion of velocity calculation). 

There are several other important parameters which determine the gravitational
wave signal from the first order phase transitions, see
e.g.~\cite{Caprini:2015zlo,Caprini:2019egz}. Parameter $\alpha$ 
characterizes the strength of the phase transition. It is the ratio
of the latent heat density (heat released during the phase transition) to
the radiation energy density    
\begin{equation}
\alpha\equiv \left(\frac{g_*\pi^2 T_{\text{perc}}^4}{30}\right)^{-1}
\eval{\left(\Delta V_{\text{eff}}-\frac{T}{4}\dv{\Delta
    V_{\text{eff}}}{T}\right)}_{T_{\text{perc}}}, 
\end{equation}
where $g_*$ is the total effective number of degrees of freedom which
are in thermal equilibrium in plasma, $\Delta V_{\text{eff}}$ is the
potential value difference between the broken and unbroken phases.
Another parameter $\beta/H_c$ characterises the (inverse) timescale of
the phase transition and reads  
\begin{equation}
\frac{\beta}{H_c} \equiv \eval{T \dv{T} \left(\frac{S_3}{T}\right)
}_{T_{\text{perc}}}\,. 
\end{equation}
We calculate $\alpha$ and $\beta/H_c$ for the selected set of
benchmark models and present them in Table~\ref{table:GWparams} along
with nucleation and percolation temperatures, $T_{\text{nuc}}$ and $T_{\text{perc}}$ as well as effective number of
degrees of freedom $g_*$. 

%%%%%%%%%%%%%%%%%%%%%%%%%%%%%%%%%%%%%%%%%%%%%%%%%%%%%%%%%
\subsection{Gravitational waves}

Calculation of spectrum of gravitational waves produced during a first-order
phase transition was a subject of numerous recent studies. Here we follow~\cite{Caprini:2015zlo,Caprini:2019egz} to estimate the GW power spectrum produced during the EWPT within the model with light sgoldstinos. For phase transitions found in this work the gravitational wave signal is mostly produced by sound waves and magnetohydrodynamic (MHD) turbulence in plasma, while the impact of released kinetic energy of bubble walls can be
neglected. The relative contribution of gravitational waves to the present energy density of the Universe reads 
\begin{equation} \label{eq:OmegaGW}
\Omega_{\text{GW}}h^2=\Omega_{\text{sw}}h^2+\Omega_{m}h^2,
\end{equation}
where the first term comes from sound waves and the second one is the MHD turbulence contribution; $h\simeq 0.7$ is the value of the Hubble constant in terms of 100\,(km/s)/Mpc. 
The general form of such signal spectra are usually parametrized as
\begin{equation}
\Omega_{\text{sw}}h^2=1.23\cdot 10^{-5}\frac{v_w H_c}{g_*^{1/3}\beta}\left(\frac{\kappa_{\text{sw}}\alpha}{1+\alpha}\right)^2 S_{\text{sw}}(f)\Upsilon,
\end{equation}
\begin{equation}
\Omega_{m}h^2=1.55\cdot 10^{-3}\frac{v_w H_c}{g_*^{1/3}\beta}\left(\frac{\kappa_{m}\alpha}{1+\alpha}\right)^{3/2} S_{m}(f),
\end{equation}
where the bubble wall velocity $v_w$ we fix to be equal 0.55 (a typical value for phase transitions), 
\begin{equation}
\kappa_{\text{sw}}=\frac{c_s^{11/5}k_a k_b}{\left(c_s^{11/5}-v_w^{11/5} \right)k_b + v_w c_s^{6/5}k_a}, \quad \kappa_m=0.05\kappa_{\text{sw}}, \quad c_s=\frac{1}{\sqrt{3}}, 
\end{equation}
\begin{equation} 
k_a=\frac{6.9 v_w^{6/5}\alpha}{1.36-0.037\sqrt{\alpha}+\alpha}, \quad k_b=\frac{\alpha^{2/5}}{0.017+(0.9997+\alpha)^{2/5}},
\end{equation}
$S_{\text{sw}}(f)$ and $S_{m}(f)$ are the spectrum shapes,
\begin{equation}
S_{\text{sw}}(f)=\left(\frac{f}{f_{\text{sw}}} \right)^3 \left(\frac{7}{4+3(f/f_{\text{sw}})^2} \right)^{7/2},  
\end{equation}
\begin{equation}
S_{m}(f)=\frac{\left(f/f_{m} \right)^3 }{\left(1+f/f_{m}\right)^{11/3}\left(1+\frac{8\pi f}{h_*}\right)},
\end{equation}
where
\begin{equation}
h_*=1.65\cdot 10^{-5}\text{ Hz}\left(\frac{T}{100\text{ GeV}} \right)\left(\frac{g_*}{100} \right)^{1/6}
\end{equation}
and peak frequencies are given by
\begin{equation} \label{eq:fpeak}
f_{\text{sw}}\simeq \frac{1.15\beta h_*}{v_w H_c}, \quad\quad f_m\simeq \frac{1.65\beta h_*}{v_w H_c}.
\end{equation}
The factor
\begin{equation}
\Upsilon=1-\frac{1}{\sqrt{1+2\tau_{sw}H_c}}\,,
\end{equation}
where
\begin{equation}
    \tau_{sw}H_c\sim (8\pi)^{1/3}\frac{v_w}{\beta/H_c~\bar{U_f}}, \quad \bar{U_f}\sim \sqrt{\frac{3\alpha}{4(1+\alpha)}\kappa_{sw}}
\end{equation}
accounts for the effective lifetime of acoustic wave as a source of gravitational waves, see refs.~\cite{Ellis:2019oqb,Guo:2020grp}. 

\begin{figure} [t]
  \begin{center}
    \includegraphics[width=0.85\textwidth]{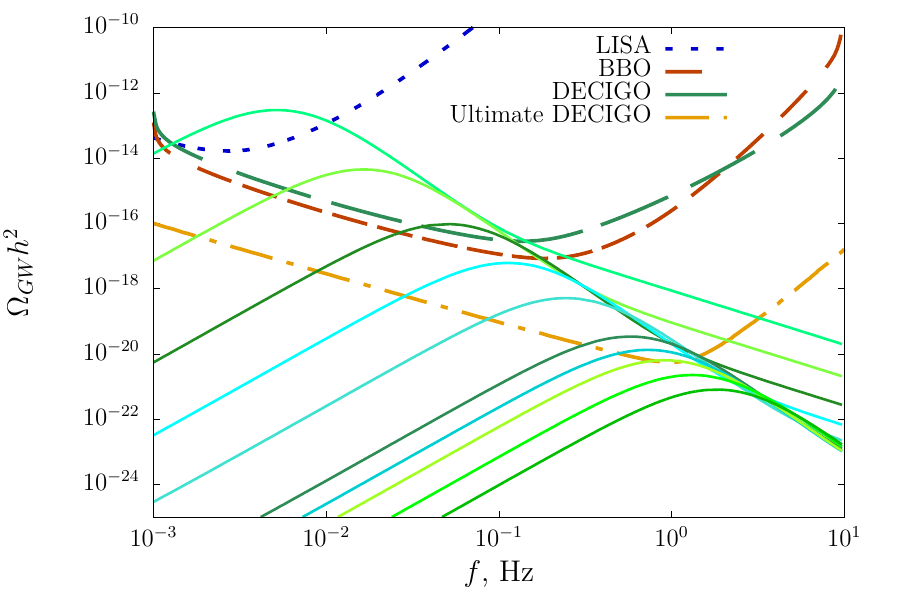}
    \caption{\label{fig:OmegaGW} Gravitational wave spectra from the first-order phase transition for points $Q_1,\,Q_2,\dots, Q_{10}$ (from top to bottom, solid lines) in model parameter space, see Table\,\ref{table:GWparams}, and sensitivity curves of LISA, BBO, DECIGO, Ultimate DECIGO experiments (dashed lines) \cite{Schmitz:2020syl,Ringwald:2020vei}.}
  \end{center}
\end{figure}

\begin{figure} [!htb]
  \begin{center}
    \includegraphics[width=\textwidth]{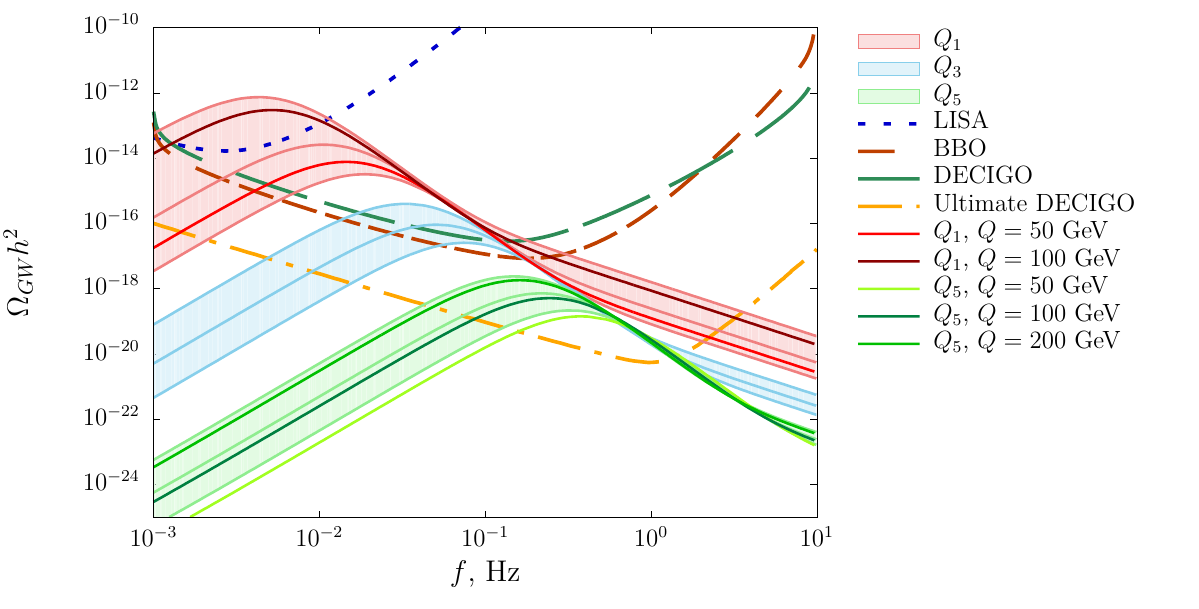}
    \caption{\label{fig:GWRenScale} Solid lines: the gravitational wave spectra from the first-order phase transition for points $Q_1$ and $Q_5$ computed adopting the renormalization scale as $Q=50$~GeV, $100$~GeV and $200$~GeV. For point $Q_1$ with $Q=200$~GeV the bubble nucleation does not take place. Filled bands: the GW spectra for points $Q_1,\,Q_3,\, Q_5$ for the variation of the renormalization scale from $2T_{perc}$ (upper edges) to $T_{perc}/2$ (bottom edges) through $T_{perc}$ (central lines).}
  \end{center}
\end{figure}

The gravitational wave spectra obtained for points $Q_1$--$Q_{10}$
using data from Table~\ref{table:GWparams} and
\eqref{eq:OmegaGW}--\eqref{eq:fpeak} are shown in
Figure~\ref{fig:OmegaGW}.  
Maxima of these signals fall into the frequency range $10^{-3}-10$~Hz. Figure~\ref{fig:OmegaGW} also shows the sensitivity curves of the
space gravitational wave observatories LISA, BBO, DECIGO and Ultimate
DECIGO. One can see that these experiments can potentially register
the predicted signals. 
The results shown in Figure~\ref{fig:OmegaGW} were obtained with renormalization scale $Q=100$~GeV. To partly estimate possible theoretical uncertainties in these predictions we study the dependence of the gravitational wave signal on our choice of the renormalization scale by varying it between $Q/2$ and $2Q$. In Figure~\ref{fig:GWRenScale} 
we show predictions for GW power spectra for models $Q_1$ and $Q_5$ obtained with different values of the renormalization scale $Q$. For benchmark model $Q_5$ the predicted signal varies within up to two orders of magnitude when calculated for $Q$ from 50 to 200~GeV. At the same time for $Q_1$ the similar difference for $Q=50$ and 100~GeV is much larger (of order $10^3$) while for $Q=200$~GeV the electroweak phase transition is found not to take place. We also link the renormalization scale to the percolation temperature and plot our results for points $Q_1,\,Q_3,\,Q_5$ as bands filled between the lines $Q=T_{perc}/2$ and $Q=2T_{perc}$ with the central line at $Q=T_{perc}$. This underlines considerable theoretical uncertainties pertinent to these predictions (see e.g.~\cite{Gould:2021oba} for a recent discussion on this subject). Table~\ref{tab:GWrenscale}
\begin{table}[h]
  \centering
  \begin{tabular}{|c|c|c|c|c|c|c|}
    \hline
    BM & Q, GeV & $T_{\text{nuc}}$, GeV & $T_{\text{perc}}$, GeV & $\alpha\cdot 10^3$ & $\beta/H_c$ & $g_*$\\
    \hline
    \multirow{3}{*}{$Q_1$} & 109.7 & 60.71 & 54.88 & 168 & 228 & 97.25\\
    \cline{2-7}
     & 68.4 & 72.59 & 68.49 & 62.5 & 442 & 107.75\\
    \cline{2-7}
     & 39.3 & 81.75 & 78.63 & 35.9 & 683 & 107.75\\
    \hline
    \multirow{3}{*}{$Q_3$} & 181.4 & 93.17 & 90.71 & 20.8 & 1040 & 107.75\\
    \cline{2-7}
     & 97.9 & 99.80 & 97.86 & 15.0 & 1490 & 107.75\\
    \cline{2-7}
     & 51.9 & 105.28 & 103.74 & 11.6 & 2060 & 107.75\\
    \hline
    \multirow{3}{*}{$Q_5$} & 231.9 & 116.94 & 115.93 & 7.06 & 3670 & 107.75\\
    \cline{2-7}
     & 120.3 & 121.02 & 120.27 & 5.77 & 5190 & 107.75\\
    \cline{2-7}
     & 62.0 & 124.53 & 123.98 & 4.80 & 7530 & 107.75\\
    \hline
  \end{tabular}
  \caption{\label{tab:GWrenscale} The renormalization scale $Q$, nucleation and percolation temperatures, parameters $\alpha$ and $\beta/H_c$ for BM $Q_1,\,Q_3,\,Q_5\,$. For each point we calculate the first order electroweak phase transition parameters taking the renormalization scale $2T_{perc}\,$, $T_{perc}$ and $T_{perc}/2$.}
\end{table}
shows the changes in the phase transition parameters crucial for the GW spectra.

\subsection{Relation to parameters of supersymmetric model}
So far we conveniently discussed physics of the electroweak phase transition
in terms of parameters entering the low-energy effective
potential~\eqref{eq:V0}. Let us relate the parameters of the benchmark
models shown in Table~\ref{table:GWparams} to the coupling constants
of the underlying supersymmetric model~\eqref{eq:1}. As we use
bottom-up approach and the corresponding higher energy theory has a lot of additional
degrees of freedom, in deducing a viable scenario we have a selection of several mass scales, such as supersymmetry breaking scale
$\sqrt{F}$ and common scale of superpartners. The latter
should be lower than $\sqrt{F}$ as we discuss in
Section~\ref{SUSYmodel}. 
The dimensionless coupling constants of the
chosen benchmark models turn out to be fairly large. Therefore,
the possibility of perturbative treatment of the model up to the supersymmetry
breaking scale requires absence of any Landau pole for these
coupling constants up to $\sqrt{F}$. We exploit the renormalization group
equations\footnote{We present the set of RGE equations for the low
energy model in appendix.} to tie the parameters at the electroweak
scale used in the study of EWPT to the parameters of the high energy
theory imposing tree level matching
conditions~\eqref{eq:coupl1}--\eqref{eq:coupl11} for an illustration.
Thus, in Table~\ref{tab:susy_parameters}
\begin{table} [h]
  \centering
  \begin{tabular}{|c||c|c|c|}
    \hline
    $\sqrt{F},$ TeV & 10 & 30 & 100\\
    \hline
    $m_A,$ TeV & 3.5 & 7.0 & 9.0\\
    \hline
    $\mu,$ TeV & -2.5 & -6.0 & -8.0\\
    \hline
    $Q,$ TeV & 4 & 8 & 10 \\
    \hline
    $\lambda_1(Q)$ & 0.0978 & 0.0962 & 0.0961\\
    \hline
    $\lambda_2(Q)$ & 1.10 & 1.18 & 1.20\\
    \hline
    $\lambda_3(Q)$ & 0.866 & 0.952 & 0.984\\
    \hline
    $\lambda_4(Q)$ & 0.0421 & 0.0472 & 0.0491\\
    \hline
    $\lambda_5(Q)$ & 0.0565 & 0.0624 & 0.0645\\
    \hline
    $\lambda_6(Q)$ & 0.0536 & 0.0641 & 0.0681\\
    \hline
    $\mu_1(Q),$ GeV & 6.78 & 7.23 & 7.39\\
    \hline
    $\mu_2(Q),$ GeV & 8.54 & 9.65 & 10.1\\
    \hline
    $\mu_3(Q),$ GeV & 6.40 & 6.81 & 6.96\\
    \hline
    $\tilde{M}_1^2(Q)\text{, GeV}^2$ & $5.15\cdot 10^3$ & $4.39\cdot 10^3$ & $4.13\cdot 10^3$\\
    \hline
    $\tilde{M}_2^2(Q)\text{, GeV}^2$ & $88.5\cdot 10^3$ & $91.0\cdot 10^3$ & $91.9\cdot 10^3$\\
    \hline
    $\tilde{M}_3^2(Q)\text{, GeV}^2$ & -$50.1\cdot 10^3$ & -$51.0\cdot 10^3$ & -$51.4\cdot 10^3$\\
    \hline
    $C_3(Q)\text{, GeV}^3$ & -$210\cdot 10^3$ & -$222\cdot 10^3$ & -$226\cdot 10^3$\\
    \hline
    $\delta_{\lambda_2}$ & 1.11 & 1.19 & 1.21\\
    \hline
    $\delta_{\lambda_3}$ & 0.866 & 0.952 & 0.984\\
    \hline
    $\delta_{\lambda_4}$ & 0.0424 & 0.0476 & 0.0496\\
    \hline
    $\delta_{\lambda_5}$ & 0.0565 & 0.0624 & 0.0645\\
    \hline
    $\delta_{\lambda_6}$ & 0.0536 & 0.0641 & 0.0681\\
    \hline
    $\delta_{\mu_1},$ GeV & -26.4 & -42.8 & -4.97\\
    \hline
    $\delta_{\mu_2},$ GeV & 6.04 & 6.83 & 7.12\\
    \hline
    $\delta_{\mu_3},$ GeV & 4.53 & 4.82 & 4.92\\
    \hline
    $\delta_{C_3}\text{, GeV}^3$ & -$149\cdot 10^3$ & -$157\cdot 10^3$ & -$160\cdot 10^3$\\
    \hline
  \end{tabular}
  \caption{\label{tab:susy_parameters} Reconstruction of physical parameters at high
    energies for the benchmark model $Q_2$ with $\tan \beta=10$. 
  } 
\end{table}
we show three exemplary scenarios corresponding to the low-energy
benchmark model $Q_2$. Namely, each column correspond to a particular
choice of $\sqrt{F}$, $m_{A}$ and $\mu$.

The direct experimental constraints on 
$\sqrt{F}$ are derived from collider searches for processes with
missing energy signature and these constraints are of order a few TeV, see
e.g.~\cite{Brignole:1997sk,Brignole:1998me,Maltoni:2015twa,Demidov:2017rmi}.
Direct searches for superpartners at ATLAS and CMS experiments \cite{Kim:2019vcp,Sirunyan:2019mbp,Sirunyan:2019hzr,Sirunyan:2018ell,Sirunyan:2018psa,Aaboud:2018zeb,Aaboud:2018doq} put model-dependent constraints on soft masses, in particular, on gluino mass. The lower bound on gluino mass in these experiments is lower than 3 TeV.
In particular, interpretation of the LHC negative results for gluinos and squarks within GMSB-like scenarios put lower bounds on maximal values of soft
SUSY breaking parameters of order~2~TeV, see e.g.~\cite{ATLAS:2018zzq}.
Therefore, we choose values of $\sqrt{F}$, $m_{A}$ and $\mu$ somewhat
above this energy scale. Table~\ref{tab:susy_parameters} shows values
of parameters of the low-energy theory at a scale $Q$ about the superpartner masses.  We have checked for each of the benchmark model that
position of the Landau pole (if any) for each coupling is well above the supersymmetry
breaking  scale, see appendix. It justifies the description of the EWPT entirely in terms of the effective field theory we adopted. As it can be valid up to the energy scale much higher than the electroweak scale, the model phenomenology at LHC can be also described in terms of the effective scalar potential and two new ingredients -- sgoldstinos. From Tables~\ref{table:GWparams} and~\ref{tab:susy_parameters} we see that
dimensionless coupling constants in the scalar sector, in particular, those responsible for interaction between sgoldstino and Higgs fields,  are of order unity.  This is a generic feature for all our benchmark points, indicating that existence of the first order EWPT requires interactions of additional scalar degrees of freedom with the Higgs boson not to be weak~\cite{Ramsey-Musolf:2019lsf}. 
Looking at the matching
conditions~\eqref{eq:coupl1}--\eqref{eq:coupl11} one concludes that for chosen values of model parameters and given the hierarchies
$m_{soft}, \mu\lsim \sqrt{F}$ and $m_{s, phys}, m_{p, phys}\ll
\sqrt{F}$ it is difficult (if not impossible) to obtain
such large values of $\lambda_2$ and $\lambda_3$ without introducing
the '$\delta$'-set of operators in the K\"{a}hler potential
in~\eqref{eq:lagr_phi}. The viable choices of values of the corresponding coupling constants
behind these operators are presented in
Table~\ref{tab:susy_parameters} for completeness.  Here we set $\tan{\beta}=10$. 

Finally, let us briefly discuss possible collider phenomenology of presented scenario associated with that part of the model, which is  responsible for  the first order electroweak phase transition. The new lightest degree of freedom, scalar sgoldstino, can show up in several signatures.  The details are defined in particular by the scalar potential~\eqref{eq:V0} describing  sgoldstino interactions with the Higgs boson. As for interactions of sgoldstino with other SM particles they are fixed by terms in the Lagrangian~\eqref{sfermion-masses},~\eqref{trilinear-terms},~\eqref{gaugino-masses} whose couplings are determined by the soft parameters. The sgoldstino direct production is naturally dominated by the gluon fusion. Its rate crucially depends on the relation between soft gluino mass $M_3$ and supersymmetry breaking scale and thus is largely model dependent. 
Prospects of such searches and bounds from existing experimental results have been discussed in several studies previously, see  e.g.~\cite{Gorbunov:2002er,Asano:2017fcp,Demidov:2017rmi,Demidov:2020jne}.
In general, sgoldstino can reveal themselves either through mixing with the Higgs boson or via triple and quartic interactions involving the Higgs boson and sgoldstino. The sgoldstino-Higgs mixing angle is determined by the mass matrix~\eqref{mmatrix} and for the found benchmark models it can be quite prominent. In particular, the sine squared of the mixing angle between scalar sgoldstino and the Higgs boson can reach values up to 0.15 for the chosen set of benchmark models. The mixing results in modification of the Higgs boson coupling constants and therefore the model can be  tested in near future by more accurate measurements of the decay and production rates of the Higgs boson.  

\begin{figure}
	\begin{center}
		\begin{subfigure}{0.45\textwidth}
			\begin{center}
				\includegraphics[width=0.5\textwidth]{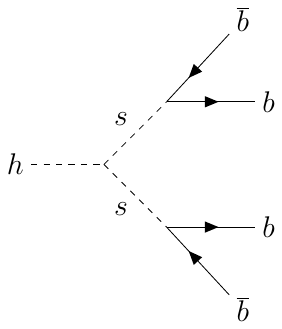}
				\caption{}
				\label{fig:h4b}
			\end{center}
		\end{subfigure}
		\begin{subfigure}{0.45\textwidth}
			\begin{center}
				\includegraphics[width=0.5\textwidth]{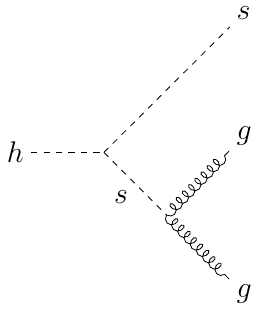}
				\caption{}
				\label{fig:h2g}
			\end{center}
		\end{subfigure}
		\caption{Possible exotic Higgs boson decays with sgoldstino involved. }
		\label{fig:hdecay}
	\end{center}
\end{figure}

Quite large trilinear and quartic couplings in addition to lightness of scalar sgoldstino in chosen benchmark models may also result in interesting phenomenology. In particular, one can expect new exotic Higgs decays to a pair of sgoldstinos, see for instance Figure\,\ref{fig:hdecay}.
If sgoldstino decays in the detector volume, the signature is four jets (Figure \ref{fig:h4b}). It is also possible to get two jets and a missing energy if one of sgoldstinos escapes the detector volume without decay (Figure \ref{fig:h2g}). 
Another interesting possibility is double scalar production which looks viable due to smallness of scalar sgoldstino mass. The low energy part of the theory is, in fact, the SM extended with complex singlet scalar and studies of double scalar production have been undertaken, e.g. in~\cite{Chen:2017qcz,Carena:2018vpt,Robens:2019kga}. It will be interesting to perform such an analysis within the model in question to study complementarity between collider and gravitational wave experiments~\cite{Alves:2018jsw,Alves:2019igs,Papaefstathiou:2020iag}.
We leave study of these and similar processes for future.

%%%%%%%%%%%%%%%%%%%%%%%%%%%%%%%%%%%%%%%%%%%%%%%%%%%%%%%%%%%%%%%%%%%%%%%%%%%%%

\section{Conclusions} \label{conclusions}
To summarise we showed the possibility of probing the sector responsible for spontaneous supersymmetry breaking using gravitational wave signal from first order EWPT. Namely, we found that in the model with low scale supersymmetry breaking whose low energy effective theory contains, apart from the SM particles, the chiral goldstino supermultiplet, there is a region in model parameter space where the EWPT can be of the first order. Using the packages \texttt{PhaseTracer} and \texttt{FindBounce} we have found several benchmark points in the model parameter space that admit the first-order EWPT at temperatures $T\approx 60-140$ GeV. For these points we have estimated the energy spectra of gravitational waves produced during the electroweak phase transition. The gravitational wave signal turns out to be of the level accessible for LISA and the proposed experiments like BBO, DECIGO and Ultimate DECIGO.  We have checked the validity of perturbative description of the scalar sector at least up to the supersymmetry breaking scale, where the model naturally must be completed with the sector responsible for the supersymmetry breaking. It justifies our description of EWPT with the effective scalar potential, which parameters can be probed at Large Hadron Collider upon performing searches for light sgoldstinos.

\section*{Acknowledgements}
The work is supported by the Russian Science Foundation RSF grant 21-12-00379. The work of EK was supported by the grant of “BASIS” Foundation no. 21-2-10-37-1.

\appendix
\section{One-loop renormalization group equations} \label{RGE}
The one-loop equations for running coupling constants $g_1$, $g_2$, $g_3$ of SM gauge groups $U(1)$, $SU(2)$, $SU(3)$ and top quark Yukawa constant $y_t$ read \cite{Luo:2002ey, Shaposhnikov:2009pv}
\begin{equation} \label{eq:betaSM}
Q \pdv{g_1}{Q} = \frac{41 g_1^3}{96\pi^2}, \quad\quad Q\pdv{g_2}{Q} = -\frac{19 g_2^3}{96\pi^2}, \quad\quad Q\pdv{g_3}{Q} = -\frac{7 g_3^3}{16\pi^2},
\end{equation}
\begin{equation}
Q \pdv{y_t}{Q} = \frac{1}{16\pi^2}\left[\frac{9}{2}y_t^3 - 8g_3^2 y_t - \frac{9}{4}g_2^2 y_t - \frac{17}{12}g_1^2y_t\right].
\end{equation}
The one-loop $\beta$-functions for dimensionless coupling constants are
\begin{multline} \label{eq:betal1}
Q \pdv{\lambda_1}{Q} = \frac{1}{16\pi^2} \left(24\lambda_1^2+\frac{1}{2}\lambda_{hs}^2+\frac{1}{2}\lambda_{hp}^2 + 12\lambda_1 y_t^2 - 3\lambda_1 g_1^2-9\lambda_1 g_2^2-6y_t^4 + \right. \\ \left. +\frac{3}{8}\left(g_1^4+3g_2^4+2g_1^2g_2^2 \right) \right), 
\end{multline}
\begin{equation}
Q \pdv{\lambda_{hs}}{Q} = \frac{1}{16\pi^2} \left(12\lambda_1 \lambda_{hs} + 6\lambda_{hs}\lambda_s + \lambda_{hp}\lambda_{sp} + 4\lambda_{hs}^2 + \lambda_{hs}\left(6y_t^2-\frac{3}{2}g_1^2-\frac{9}{2}g_2^2 \right) \right), 
\end{equation}
\begin{equation}
Q \pdv{\lambda_{hp}}{Q} = \frac{1}{16\pi^2} \left(12\lambda_1 \lambda_{hp} + 6\lambda_{hp}\lambda_p + \lambda_{hs}\lambda_{sp} + 4\lambda_{hp}^2 + \lambda_{hp}\left(6y_t^2-\frac{3}{2}g_1^2-\frac{9}{2}g_2^2 \right)\right), 
\end{equation}
\begin{equation}
Q \pdv{\lambda_s}{Q} = \frac{1}{16\pi^2} \left(18\lambda_s^2 + 2\lambda_{hs}^2 + \frac{1}{2}\lambda_{sp}^2 \right), 
\end{equation}
\begin{equation}
Q \pdv{\lambda_p}{Q} = \frac{1}{16\pi^2} \left(18\lambda_p^2 + 2\lambda_{hp}^2 + \frac{1}{2}\lambda_{sp}^2 \right),
\end{equation}
\begin{equation}
Q \pdv{\lambda_{sp}}{Q} = \frac{1}{16\pi^2} \left(6\lambda_s \lambda_{sp} + 6\lambda_p\lambda_{sp} + 4\lambda_{hs}\lambda_{hp} + 4\lambda_{sp}^2 \right). 
\end{equation}
This result is in agreement with the renormalization group equations in \cite{Kainulainen:2015sva}, where the model with a Higgs boson and one new scalar is studied.

It is known that in the SM at high energies $Q\sim 10^{10}\text{ GeV}$ the Higgs boson self-coupling constant becomes negative \cite{Degrassi:2012ry}. The SM calculation of its evolution in the one-loop approximation gives the same scale $Q$ value about $10^8\text{ GeV}$ (see figure 8 in \cite{Brod:2020lhd}). Let us note that the second and third terms in \eqref{eq:betal1} (diagrams with scalar and pseudoscalar sgoldstinos in the loop) increase $\lambda_1$ and keep its value positive at higher energies up to the scale of Landau pole. For phase transitions considered in this paper the coupling constants have pole at the scales of order $10^8-10^9\text{ GeV}$. We expect that deriving renormalization group equations in two loops and taking into account the input of superpartners since the $10\text{ TeV}$ scale moves the pole to higher energies. 

The equations for coupling constants $\mu_i$ and mass parameters $\widetilde{M}_1^2$, $M_s^2$, $M_p^2$ read
\begin{equation}
Q \pdv{\mu_1}{Q} = \frac{1}{16\pi^2}\left(12\lambda_1\mu_1 + 4\lambda_{hs}\mu_1 + \lambda_{hs}\mu_s + \lambda_{hp}\mu_{sp} + \mu_1\left(6y_t^2 - \frac{3}{2}g_1^2 - \frac{9}{2}g_2^2\right)  \right), 
\end{equation}
\begin{equation}
Q \pdv{\mu_s}{Q} = \frac{1}{16\pi^2}\left(18\lambda_s\mu_s + 3\lambda_{sp}\mu_{sp} + 12\lambda_{hs}\mu_1 \right), 
\end{equation}
\begin{equation}
Q \pdv{\mu_{sp}}{Q} = \frac{1}{16\pi^2}\left(6\lambda_p\mu_{sp} + 4\lambda_{sp}\mu_{sp} + \lambda_{sp}\mu_s + 4\lambda_{hp}\mu_1 \right). 
\end{equation}
\begin{equation}
Q \pdv{\widetilde{M}_1^2}{Q} = \frac{1}{16\pi^2}\left(12\lambda_1 \widetilde{M}_1^2 - \lambda_{hs}M_s^2 - \lambda_{hp}M_p^2 - 2\mu_1^2 + \widetilde{M}_1^2\left(6y_t^2 - \frac{3}{2}g_1^2 - \frac{9}{2}g_2^2\right) \right), 
\end{equation}
\begin{equation}
Q \pdv{M_s^2}{Q} = \frac{1}{16\pi^2}\left(6\lambda_s M_s^2 + \lambda_{sp}M_p^2 - 4\lambda_{hs}\widetilde{M}_1^2 + \mu_s^2 + \mu_{sp}^2 + 4\mu_1^2 \right), 
\end{equation}
\begin{equation} \label{eq:RGEMp}
Q \pdv{M_p^2}{Q} = \frac{1}{16\pi^2}\left(6\lambda_p M_p^2 + \lambda_{sp}M_s^2 - 4\lambda_{hp}\widetilde{M}_1^2 + 2\mu_{sp}^2 \right). 
\end{equation}
These equations can be verified by comparison with one-loop $\beta$-functions and anomalous dimensions in \cite{Holthausen:2011aa}, \cite{Ghorbani:2021rgs}.

\end{document}